# Precise Clock Synchronization in the Readout Electronics of WCDA in LHAASO

Lei Zhao, *Member, IEEE,* Shaoping Chu, Cong Ma, Xingshun Gao, Yunfan Yang, Shubin Liu, *Member, IEEE*, and Qi An, *Member, IEEE*

***Abstract*—The Water Cherenkov Detector Array (WCDA) is one of the key parts in the Large High Altitude Air Shower Observatory (LHAASO). In the WCDA, 3600 Photomultiplier Tubes (PMTs) and the Front End Electronics (FEEs) are scattered within a 90000 m$^2$ area, while a time measurement resolution better than 0.5 ns is required in the readout electronics. To achieve such time measurement precision, high quality clock distribution and synchronization among the 400 FEEs (each FEE for 9 PMTs readout) is required. To simplify the electronics system architecture, data, commands, and clock are transmitted simultaneously through fibers over a 400-meter distance between FEEs and the Clock and Data Transfer Modules (CDTMs). In this article, we propose a new method based on the White Rabbit (WR) to achieve completely automatic clock phase alignment between different FEEs. The original WR is enhanced to overcome the clock delay fluctuations due to ambient temperature variations. This paper presents the general scheme, the design of prototype electronics, and initial test results. These indicate that a clock synchronization precision better than 50 ps is achieved over 1 km fibers, which is well beyond the application requirement.**

## I. INTRODUCTION

THE Water Cherenkov Detector Array (WCDA) [1] is one of the key parts in the Large High Altitude Air Shower Observatory (LHAASO), which is in the R&D phase [2]. It consists of four 150 m ×150 m water ponds. In each pond there are 900 Photomultiplier Tubes (PMTs) placed under water, and thus a total of 3600 PMTs need to be read out. The signal from a PMT has a dynamic range from 1 to 4000 Photo Electrons (P.E.). Both high precision charge (required resolution: 30%@1 P.E. and 3%@4000 P.E.) and time (required resolution: bin size <1 ns and RMS <0.5 ns) measurements are required over this range [3]. To guarantee a good resolution, the Front End Electronics are placed close to the PMTs. These FEEs are responsible for signal amplification, shaping, analog-to-digital conversion, and time discrimination, as well as time and charge calculation based on FPGA devices. The output data from the FEEs are transmitted over a distance of up to 400 meters to the Clock and Data Transfer Modules (CDTMs) which further transmit the data to the Data Acquisition system (DAQ), as shown in Fig. 1. The time measurement results of different PMT channels are correlated to build events. Therefore, to achieve a good time resolution, a high quality clock has to be distributed over this large area, and the clock phases among different FEE nodes must be aligned (i.e. compensated) with good precision.

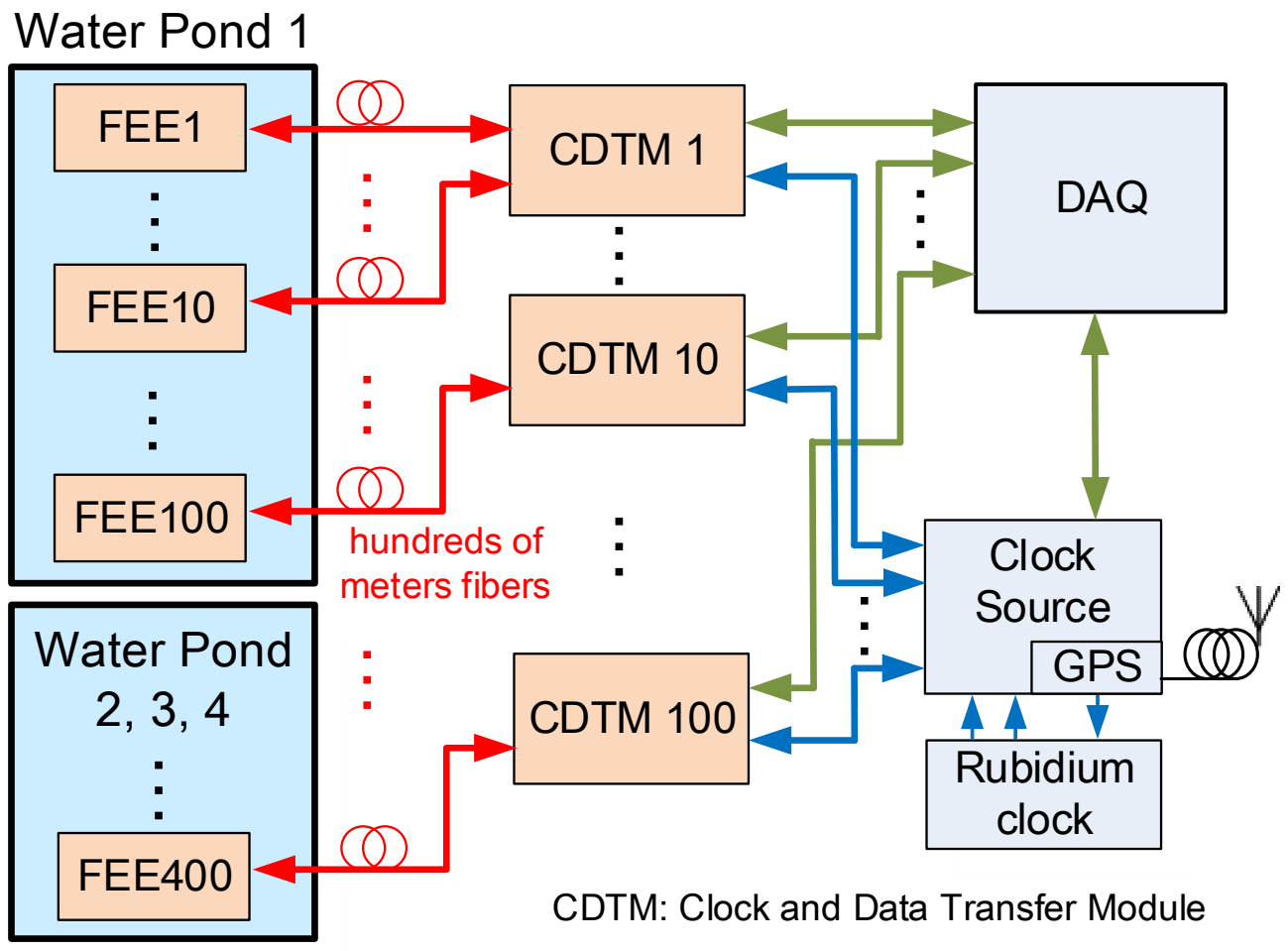

Fig. 1. Architecture of the readout electronics system for the WCDA in LHAASO.

In a traditional clock distribution system, clock signals are distributed via dedicated paths, e.g. through coaxial cables or optical fibers, separated from the data transmission path. Since fiber based transmission can effectively mitigate Electromagnetic Interference (EMI) and isolate the ground connection over long distance, it is often employed in applications that require high fidelity clock distribution [4], [5]. However, in experiments with detectors scattered over a large area, high system complexity is inevitable with this method. To solve this problem, several transmission methods based on modulation techniques have been proposed [6]–[9] to transport clock and data simultaneously over the same media.

Any propagation delay difference on these long transmission

Manuscript received April 29, 2015; revised September 16, 2015; accepted September 29, 2015. This work was supported in part by the National Natural Science Foundation of China under Grant 11175174, in part by the Knowledge Innovation Program of the Chinese Academy of Sciences under Grant KJCX2-YW-N27, and in part by the CAS Center for Excellence in Particle Physics (CCEPP).

The authors are with the State Key Laboratory of Particle Detection and Electronics, University of Science and Technology of China, Hefei, 230026; and Modern Physics Department, University of Science and Technology of China, Hefei, 230026, China (telephone: 086-0551-63601925, corresponding author: Qi An, e-mail: anqi@ustc.edu.cn).

paths will cause a phase alignment mismatch between the different FEEs. Usually, static differences are compensated offline. However, ambient temperature variations will cause dynamic fluctuations of this delay difference. To address this issue, in the BELLE experiment at KEKB [10] and BESIII TOF electronics [5], Phase Stabilized Optical Fibers (PSOFs) are employed for clock distribution. For example, BESIII TOF employs fibers with a temperature coefficient as low as 0.4 ppm/°C. However, the cost of this kind of product seems prohibitive in large scale experiments. In the ANTARES (Astronomy with a Neutrino Telescope and Abyss environmental REsearch) experiment, Time-to-Digital Converter (TDC) modules are used to measure the clock propagation delay variation with a precision of 0.5 ns, and then the clock phases can be calibrated and aligned [6]. In the CNGS (CERN Neutrino beam to Gran Sasso) experiment [11], the White Rabbit (WR) method [12] proposed by CERN based on the Precise Timing Protocol (PTP) [13] is employed. This technique can achieve sub-nanosecond clock phase compensation. In 2013, based on the principle of WR, we achieved a phase compensation precision of less than 100 ps when varying the temperature of optical fibers from 30 °C to 75 °C [14]. In 2014, Li, *et al.* obtained a precision of 150 ps (50 ps standard deviation) when applying temperature changes from -10 °C to 55 °C to the electronics and optical fibers. This method uses thermometers to achieve real-time temperature correction [15].

In the WCDA readout electronics of LHAASO, fiber based transmission is employed to transmit clock, data, and commands together to simplify the system architecture. A good clock compensation method has to be proposed to obtain a good time measurement resolution among all the 3600 PMT channels. To achieve a good clock phase compensation of less than 100 ps and completely real-time correction, we proposed a new method based on the basic idea of the WR technique to automatically compensate the delay increment of the upwards (from FEE to CDTM) and downwards (from CDTM to FEE) directions, without using thermometers to monitor ambient temperature. We also conducted tests to evaluate the performance of this method, as described in Section IV.

## II. Methods and Implementation

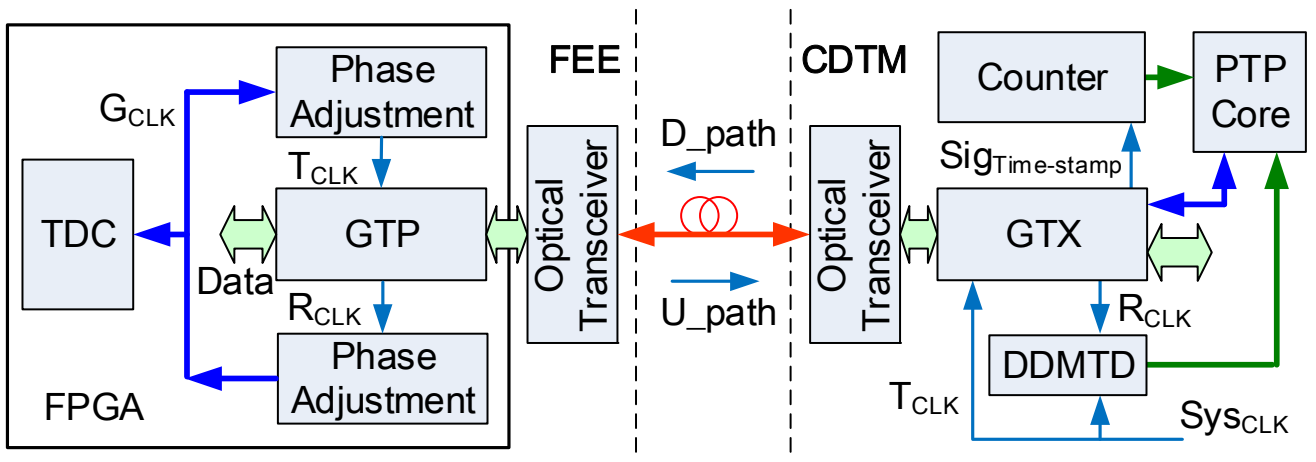

Fig. 2. Clock synchronization scheme.

Real-time clock phase compensation is based on the real-time delay measurement and adjustment. As shown in Fig. 2, the CDTM receives the clock “$Sys_{CLK}$” from the clock source, and distributes it to the FEE through an optical fiber. As aforementioned, since the data of the FEE are required to be accumulated and transmitted, we would like to combine the transmission of data (from FEE to DAQ), clock, and commands (from DAQ to FEE) on the same fiber to reduce the system complexity. We achieve this by using the GTP/GTX interface [16] within FPGA devices. As shown in Fig. 2, the GTP/GTX converts the data (or commands) and clock signal to high-speed serial streams, which are transformed to optical signals through the optical transceiver. In the other direction, the serial stream is converted back to recover the clock and data. Considering the recovered clock uncertainty [17], [18], “Bitslide” operation [18] is conducted by the logic in the FPGA to determine the clock phase.

We employ single-mode fibers with different wavelengths used for bi-directional transmission. We mark these two directions as “D_path” (from CDTM to FEE) and “U_path” (from FEE to CDTM) in Fig. 2. With this structure, the clock is transmitted through the fiber to FEE as the clock (“$G_{CLK}$”) of Time-to-Digital Converter (TDC) after phase adjustment. “$G_{CLK}$” is sent back through the fiber to the CDTM and recovered by the GTX in the CDTM as “$R_{CLK}$”. By real-time measurement of the phase difference between “$R_{CLK}$” and “$Sys_{CLK}$” based on the Digital Dual Mixer Time Difference measurement method (DDMTD) [19], the roundtrip delay variation of “D_path” plus “U_path” can be obtained. The measurement results are sent to the FEE for phase adjustment using a special data frame.

One problem associated with phase compensation is that the exact ratio of the "D_path" to "U_path" delays may be susceptible to ambient temperature variation. The most effective method proposed to date is to measure this ratio at different temperatures and record these values in a Look-Up-Table (LUT). Using a thermometer and the information retrieved from this LUT, real-time clock phase compensation can be made. In this paper, we proposed a new method for real-time clock phase compensation to simplify the system design, without the need of temperature monitoring with thermometers or complex LUTs.

In the first step, we need to analyze the delays in the round trip.

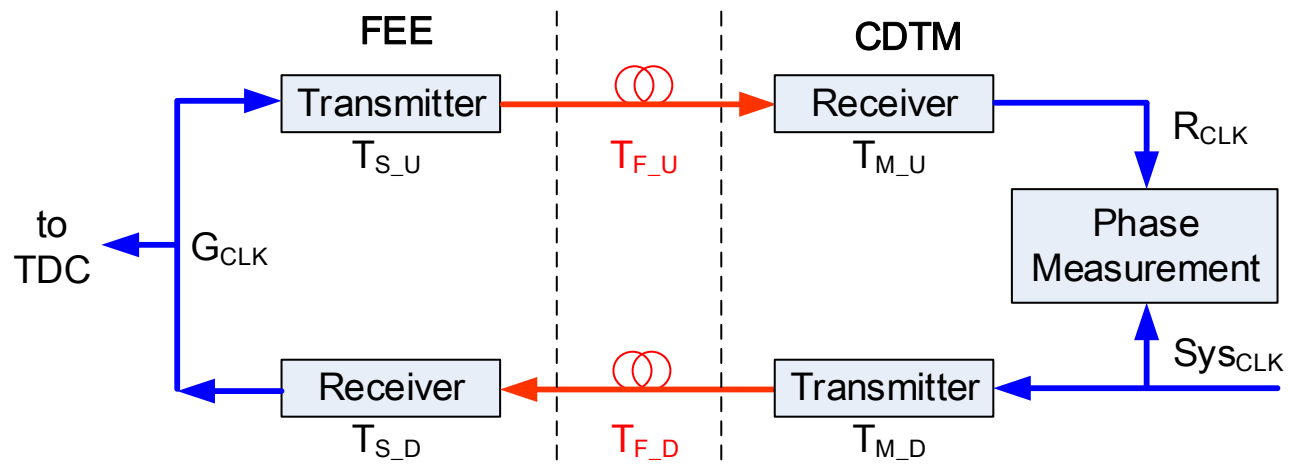

Fig. 3. Link delay model of the “D_path” and “U_path” in the loop.

As shown in Fig. 3, the roundtrip delay is measured by comparing “$R_{CLK}$” and “$Sys_{CLK}$”. This delay is actually contributed by the circuits in the FEE and CDTM, as well as the fibers. This roundtrip delay consists of two parts, as in [20]

$$T_{loop} = T_{U_PATH} + T_{D_PATH}$$
$$T_{U_PATH} = T_{S_U} + T_{F_U} + T_{M_U} \quad (1)$$
$$T_{D_PATH} = T_{S_D} + T_{F_D} + T_{M_D}$$

where $T_{loop}$ refers to the roundtrip delay; $T_{U_PATH}$ and $T_{D_PATH}$ refer to the one-way delays from FEE to CDTM and vice versa; $T_{S_U}$ and $T_{S_D}$ refer to the delays contributed by the circuits of FEE in $T_{U_PATH}$ and $T_{D_PATH}$; $T_{M_U}$ and $T_{M_D}$ refer to the delays contributed by the CDTM in the two directions; $T_{F_U}$ and $T_{F_D}$ refer to the delays contributed by the fibers.

To correctly align the clock signals received at different FEE nodes, we need to obtain the value of $T_{D_PATH}$, or the ratio of $T_{D_PATH}$ to $T_{U_PATH}$, as shown in

$$\alpha_F = \frac{T_{F_D}}{T_{F_U}}, \ \alpha_E = \frac{T_{S_D}}{T_{S_U}} \quad (2)$$

where $\alpha_F$ and $\alpha_E$ refer to the delay ratios introduced by the circuits and the fiber respectively, and these two ratios could be different. Considering CDTMs are located in the control room with a stable temperature, the variation of $T_{M_U}$ and $T_{M_D}$ can be ignored.

Efforts were devoted to measure the ratio of $T_{D_PATH}$ to $T_{U_PATH}$ in previous research [15] and it has been proven to be temperature dependent.

In the clock synchronization of WCDA readout electronics in LHAASO, the clock phase compensation is achieved by a coarse clock phase adjustment combined with a fine clock phase compensation. The former is based on the WR PTP protocol to achieve a compensation precision of one clock period (16 ns). As for the latter (fine clock phase compensation), we come up with a new method in this paper: instead of studying the total delay value of $T_{D_PATH}$ or $T_{U_PATH}$, we focus on the increment value of the delay compared to a reference when the environment temperature changes. This reference could be the delay test results at a specific ambient temperature, e.g. 22° C. The delay value of $T_{D_PATH}$ or $T_{U_PATH}$ at this temperature can be easily calibrated, which will be discussed later.

We mark the increment values as $\Delta T_{F_U}$, $\Delta T_{F_D}$, $\Delta T_{S_U}$, and $\Delta T_{S_D}$, and new ratios can be defined, as in

$$\beta_F = \frac{\Delta T_{F_D}}{\Delta T_{F_U}}, \ \beta_E = \frac{\Delta T_{S_D}}{\Delta T_{S_U}} \quad (3).$$

Now the total increment of the roundtrip delay can be expressed as

$$\begin{aligned}\Delta T_{loop} &= \Delta T_{U_PATH} + \Delta T_{D_PATH} \\ &= \Delta T_{S_U} + \Delta T_{F_U} + \Delta T_{S_D} + \Delta T_{F_D} \quad (4). \\ &= \left(1+\frac{1}{\beta_F}\right)\Delta T_{F_D} + \left(1+\frac{1}{\beta_E}\right)\Delta T_{S_D}\end{aligned}$$

$\beta_F$, $\beta_E$, and the ratio of $\Delta T_{S_D}$ to $\Delta T_{F_D}$ versus $\Delta T_{loop}$ at different temperatures can be calibrated. In real applications, we measure $\Delta T_{loop}$ in real time, and the one-way delay increment of $\Delta T_{D_PATH}$ can be determined with the ratios obtained in the calibration, without the need to monitor the ambient temperature.

In our research, we implement the electronics with an approximately symmetrical structure, including the logic architecture in the FPGA, the routing of the Printed Circuit Board (PCB), etc. By employing two of the PLLs embedded in the Xilinx Artix-7 FPGA used on the FEE, the clock phase can be adjusted with a step size as small as 15 ps, without using external circuits. As mentioned above, we also employ a single-mode fiber for bi-directional communication on one single fiber, and different wavelengths are used in the "D_path" and "U_path". Therefore, a good system symmetry and simplicity is guaranteed, and we can expect a simplified form of (4), which is actually achieved according to the results in Section III.

To simplify the analysis, we study the delays introduced by the fiber ($(1+1/\beta_F)\bullet\Delta T_{F_D}$, the first term in (4)) and by the circuits ($(1+1/\beta_E)\bullet\Delta T_{S_D}$, the second term in (4)), respectively.

## III. Delay Analysis

### A. Delay Introduced by the Fiber

First, we study the delay introduced by the fiber at different temperatures, while the electronics are placed in a temperature stable environment. To apply our method, we need to calibrate two parameters versus temperature: one is the total delay increment, as in

$$\begin{aligned}\Delta T_{F_total} &= \Delta T_{F_U} + \Delta T_{F_D} \\ &= \left(1+\frac{1}{\beta_F}\right)\Delta T_{F_D} \quad (5)\end{aligned}$$

, and the other is $\beta_F$ defined in (3).

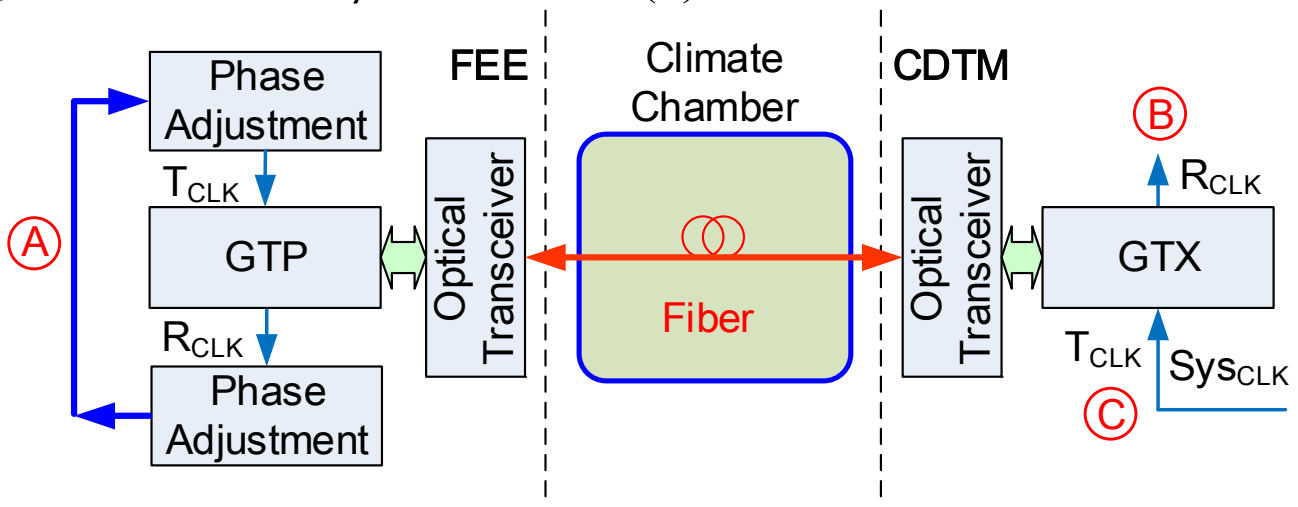

Fig. 4. Fiber delay test scheme.

We connected the three probes "A", "B", and "C" in Fig. 4 to test points of the FPGAs in the FEE and CDTM, and used a high speed oscilloscope to measure the skew between them.

In the tests to evaluate the "D_path" and "U_path" delays contributed by the fiber at different temperatures, we did not measure the clock skew directly, since the value would be limited within 16 ns. Instead we generated a pulse which is aligned with the clock at "C" in the CDTM (in Fig. 4) and transmitted it through "D_path" to the FEE. We can detect the pulse received at the FEE, which is aligned with the clock at "A". By measuring the skews between these two pulses, the "D_path" delay of fiber can be evaluated. As for the "U_path", the test method is similar. Since only the fiber is placed in the climate chamber, the electronics does not contribute to $\Delta T_{D_PATH}$ (variation of skew between "A" and "C") and $\Delta T_{U_PATH}$ (variation of skew between "B" and "A"), i.e. $\Delta T_{D_PATH} = \Delta T_{F_D}$, $\Delta T_{U_PATH} = \Delta T_{F_U}$.

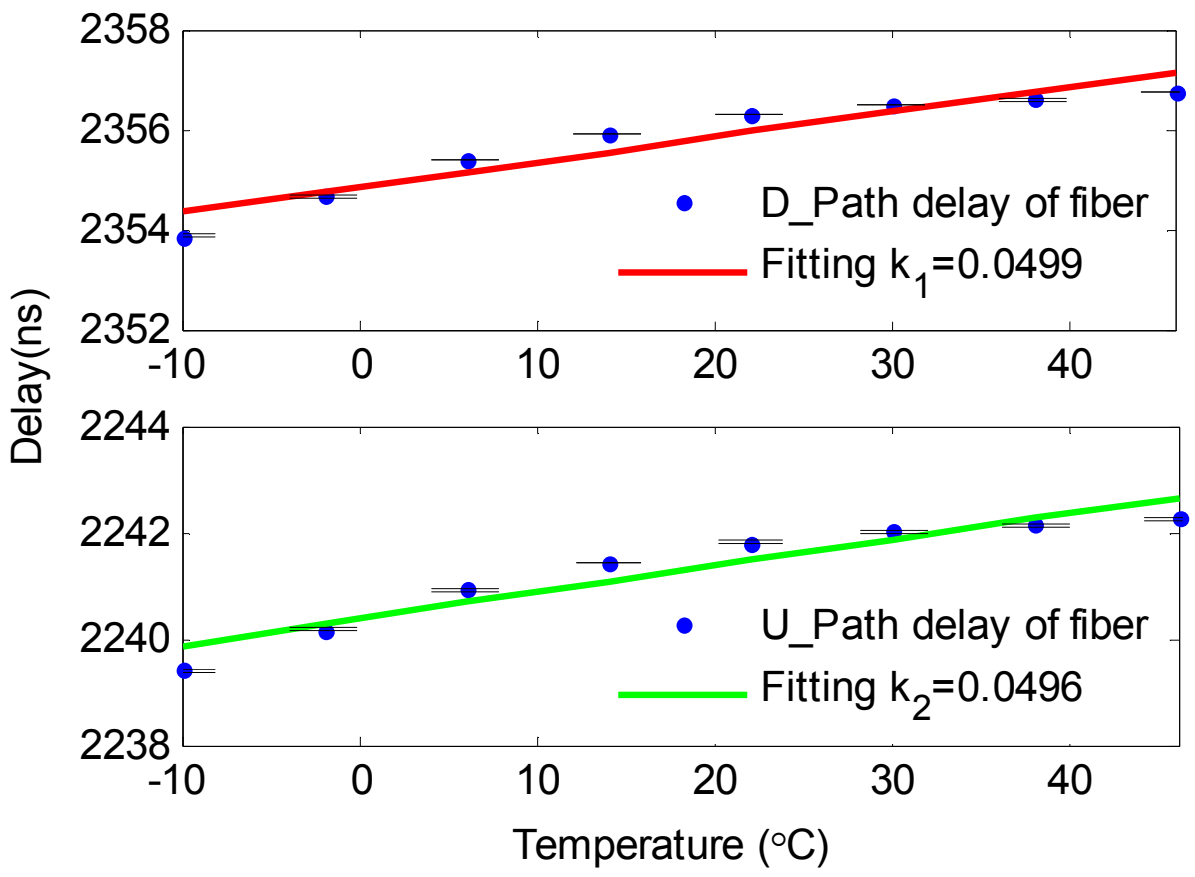

Fig. 5. Fiber delay test results (error bar is plotted with the RMS value).

Fig. 5 shows the delay measured on a 400 m fiber (type G652.D) when its temperature is changed from -10°C to 46°C. $T_{F_D}$ increases from 2353.866 ns to 2356.734 ns, while $T_{F_U}$ increases from 2239.402 ns to 2242.252 ns, which means that a time error around 3 ns would occur without clock phase alignment. Fig. 5 also indicates that $\Delta T_{F_D}$ and $\Delta T_{F_U}$ are in linear relationship with temperature approximately. From linear curve fitting, the slope ratio $k_1/k_2$ is 0.0499/0.0496. When the temperature changes to any value in this range, with the measurement result of $\Delta T_{F_total}$, $\Delta T_{F_D}$ and $\Delta T_{F_U}$ can be determined according to $k_1/k_2$, without the need to know the temperature itself. Since $k_1/k_2$ is close to 1/1, we can also try to divide $\Delta T_{F_total}$ by 2 (i.e. $\beta_F$ =1) to obtain $\Delta T_{F_D}$ and $\Delta T_{F_U}$.

Fig. 6 shows the time error results with $k_1/k_2$ *of* 0.0499/0.0496 and 1/1, respectively; the errors are both within ±25 ps and the difference is small enough for our application. If the temperature range changes, tests should be conducted to obtain the value of $k_1/k_2$. As for our application, the temperature range from -10 °C to 46 °C is large enough.

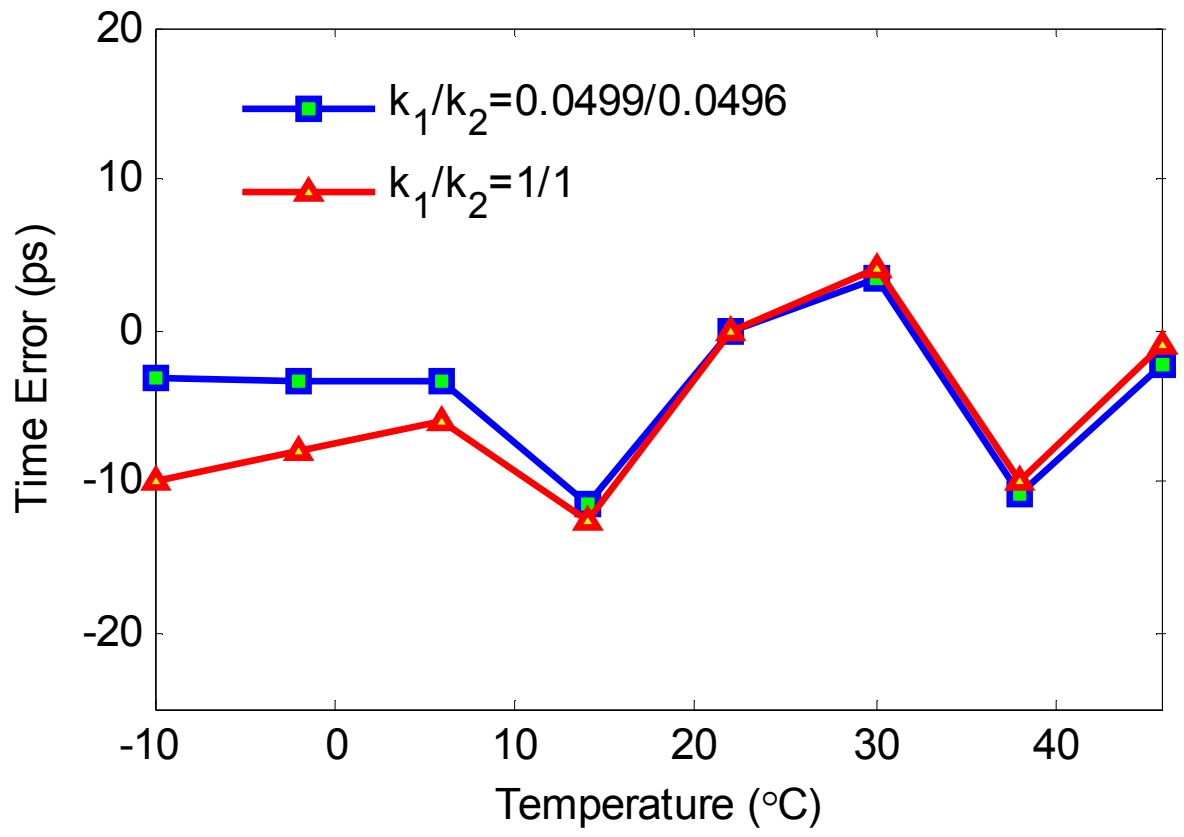

Fig. 6. Time errors with two sets of $k_1/k_2$ values.

## B. *Delay Introduced by the Circuits*

As aforementioned, we implement a symmetrical circuit structure to reduce the difference of the delay in "U_PATH" and "D_PATH" introduced by the circuits. We also conducted tests to study the delay of the electronics circuits.

The basic idea is to change the ambient temperature of the FEE, and measure the delay increment. Since the FEE is running with a changing temperature, it is no longer adequate to perform measurements at test points "A", "B", and "C". This is because the paths connecting the test points with the internal nodes in the FPGA of the FEE would also vary with temperature. In this paper, we propose another test method, as shown in Fig. 7.

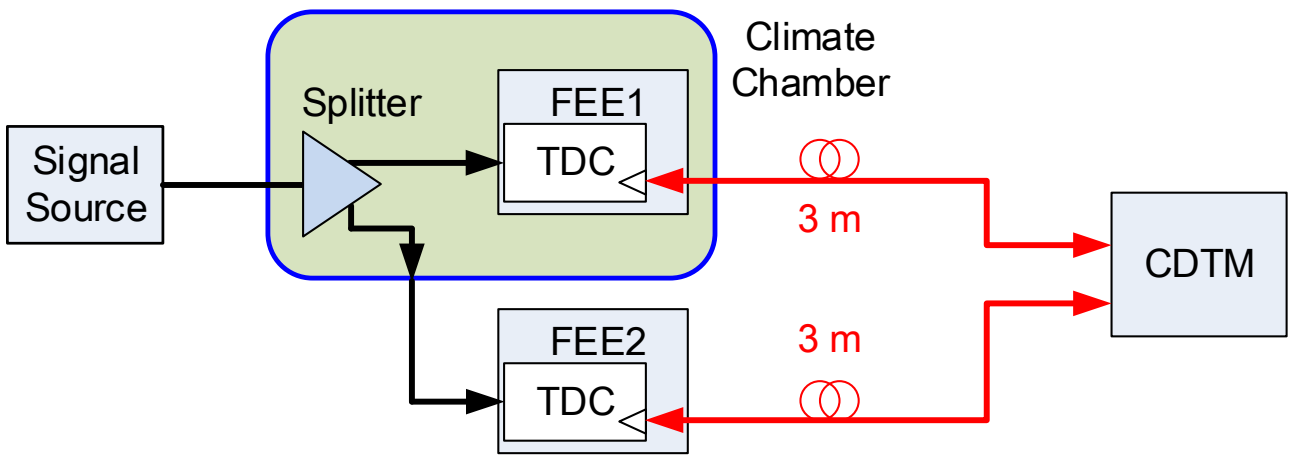

Fig. 7. Circuits delay test scheme.

We used a programmable signal generator (Tektronix model AFG3252) to produce pulses similar to that of a PMT (Hamamatsu model R5912) and fed these at the inputs of two FEEs. Each FEE contains an FPGA TDC based on the multi-phase clock interpolation technique [21]. The time of the signal leading edge is digitized by the TDC and read out to the CDTM. Then the time difference between the two TDCs' output results can be calculated by subtraction. Based on statistical analysis of a large amount of data, a precise time difference value can be obtained. The variations of this time difference reflect the change of clock phases at the level of the FEEs. We placed one of the FEEs (marked as "FEE1" in Fig. 7) in the climate chamber. Since the clock delay directly influences the time measurement results of the TDC, we can observe the delay compensation effect by tuning the value of $\beta_E$. In this test scheme, no other additional test point signal path would introduce test errors.

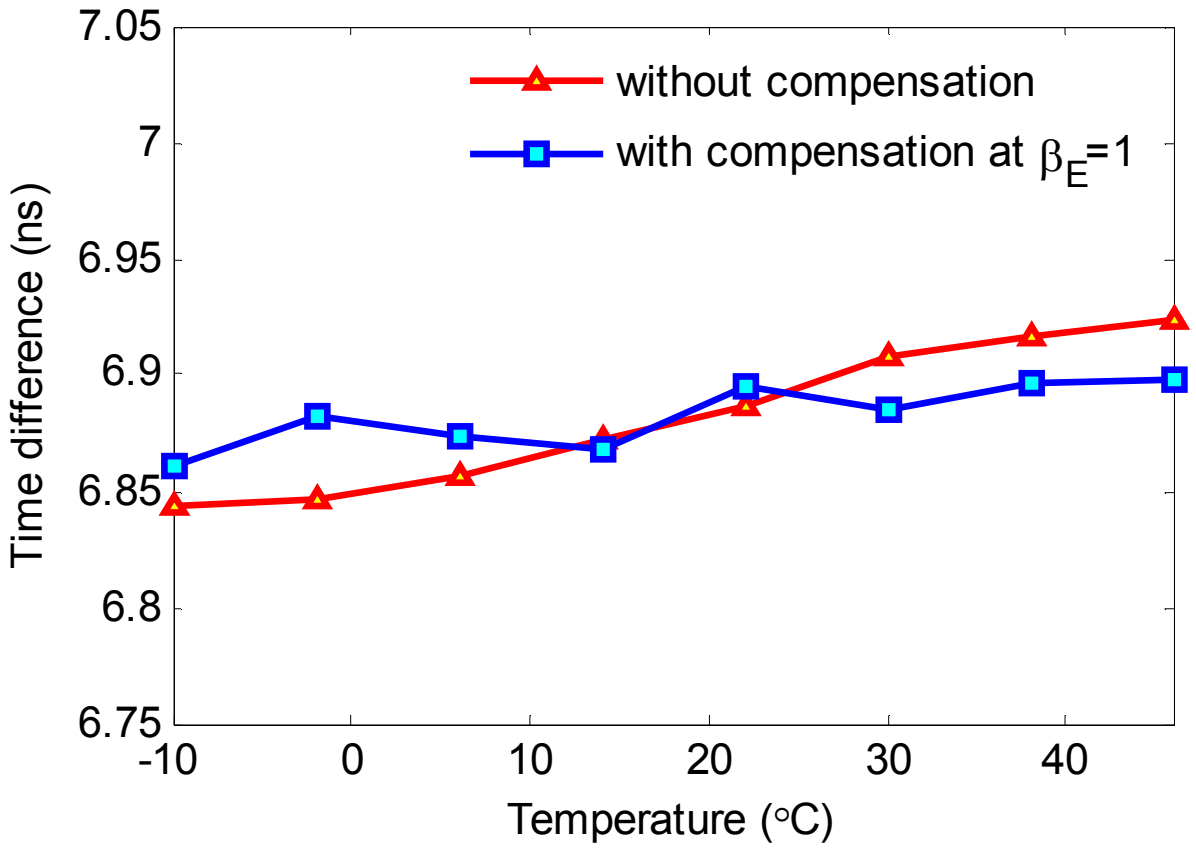

Fig. 8. Circuits delay test results.

The measured time difference between FEE1 and FEE2 without compensation is shown in Fig. 8. It can be observed that the time difference changes with temperature, which indicates fluctuation of the clock delay due to temperature drift.

As mentioned above, we can optimize the delay compensation effect by tuning $\beta_E$. Considering the symmetrical structure of the circuits, a straight forward idea is to start with $\beta_E$=1. The time measurement results with the delay compensation at $\beta_E$=1 are also shown in Fig. 8. The result indicates that the clock delay is within 40 ps, which is good enough.

## C. Compensation Method

With the above analysis and test results, $\Delta T_{loop}$ in (4) can be simplified as in

$$\begin{aligned}\Delta T_{loop} &= \Delta T_{U_PATH} + \Delta T_{D_PATH} \\ &= 2\Delta T_{F_D} + 2\Delta T_{S_D} \qquad (6), \\ &= 2\Delta T_{D_PATH}\end{aligned}$$

which means that we can directly divide $\Delta T_{loop}$ by 2 to obtain the one-way delay increment of $\Delta T_{D_PATH}$. This is rather easy to implement in real application, without the need of real temperature monitoring or complex LUTs. Even if $\beta_F \neq \beta_E$ (it is quite possible in other situations with different electronics or fiber type), the compensation method proposed in this paper still works. In the readout electronics of WCDA in LHAASO, the above tests will be conducted on the 400 FEEs. If any different values of $\beta_F$ or $\beta_E$ is found, real-time compensation is implemented according to (4), with the calibration results of $\beta_F$, $\beta_E$, and the ratio of $\Delta T_{S_D}$ to $\Delta T_{F_D}$ versus $\Delta T_{loop}$.

# IV. Evaluation and Tests

To confirm the validity of this compensation method, we conducted a series of tests in three steps. The system under test is shown in Fig. 9.

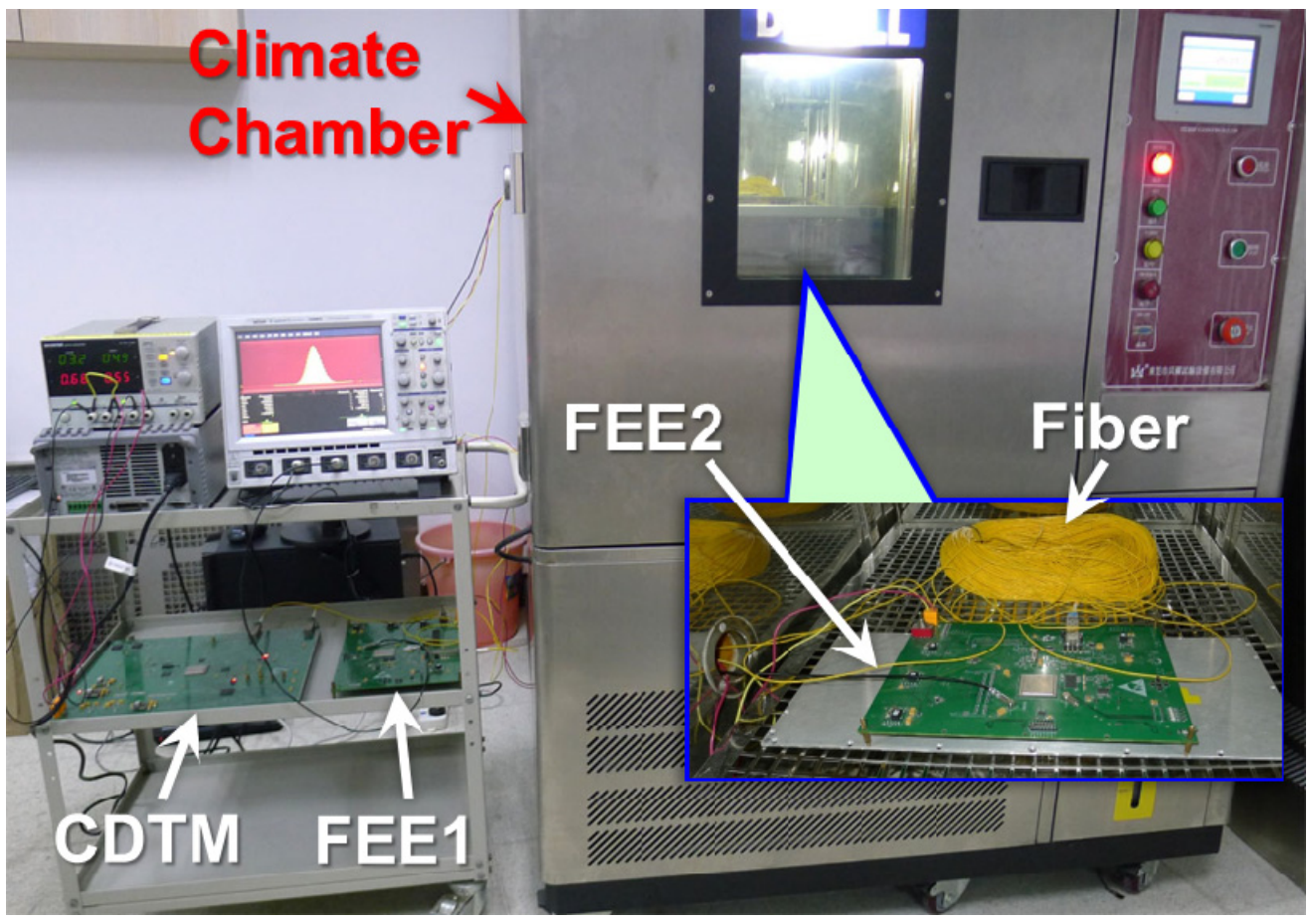

Fig. 9. System under test.

## A. Combination Tests with FEEs and Fibers with Varying Ambient Temperature

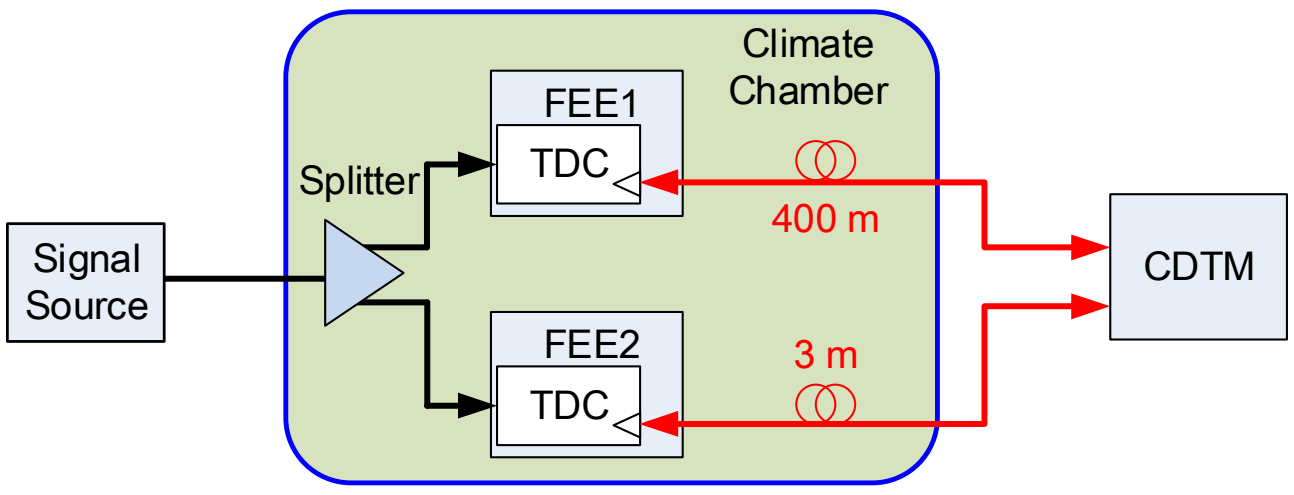

Fig. 10. Test scheme A: FEEs and Fibers all placed in the climate chamber.

In the real application, all FEEs are placed above water of the WCDA within one large hall, and connected to the CDTMs through fibers up to 400 meters long. In the first step of our evaluation, we placed both the FEEs and fibers in the climate chamber to approximate the application situation, as shown in Fig. 10. Instead of equal length, the two fibers are 400 meters and 3 meters long, respectively, in order to evaluate the compensation effect in a more severe situation. We changed the temperature inside the climate chamber from -10 °C to 46 °C to cover a sufficiently wide temperature range.

As shown in Fig. 11, the difference between the two FEE time measurement results is greatly reduced by this compensation method. The time difference result is very stable, with a maximum variation of 20 ps, which directly corresponds to the clock compensation effect.

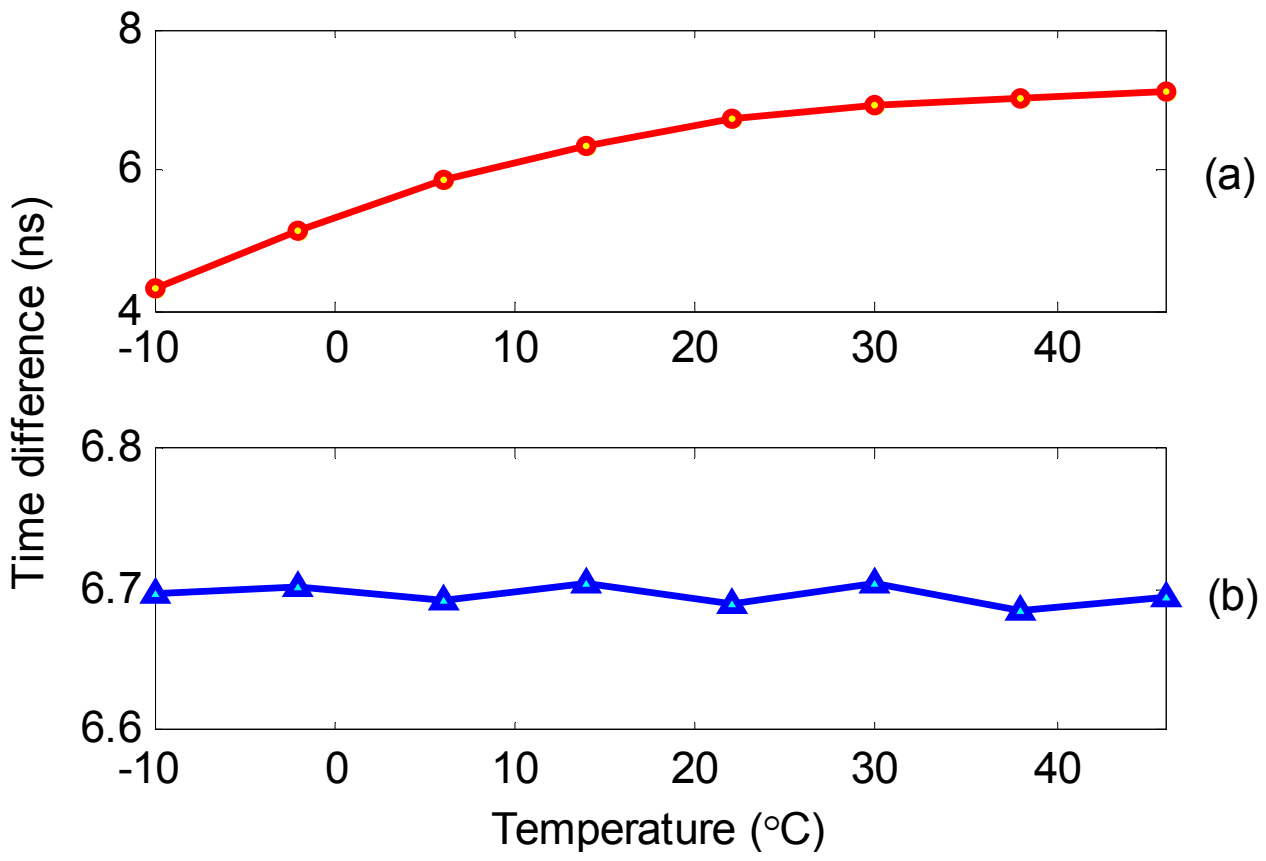

Fig. 11. Test results of scheme A. (a) without compensation; (b) with compensation.

## B. Tests with one FEE and a 400 meter Fiber with Varying Ambient Temperature

To further evaluate the effect of this compensation method in a much more severe condition, we conducted tests with scheme B, in which only one FEE and the 400 m fiber were placed in the climate chamber.

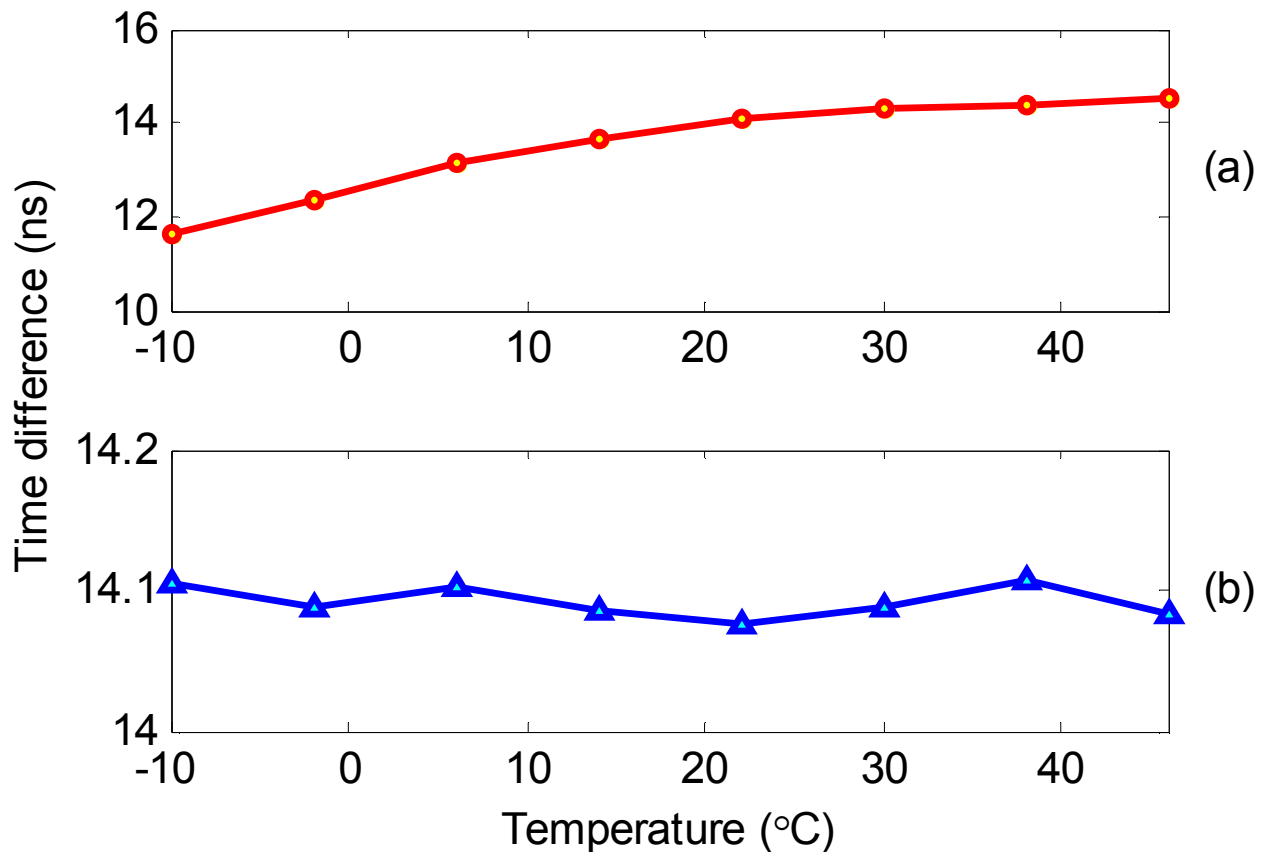

Fig. 12. Test results of scheme B. (a) without compensation; (b) with compensation.

As shown in Fig. 12, even in test scheme B, the variation of the time measurement results is within 40 ps, which is good enough and well beyond application requirement.

### *C. Tests with one FEE and a 1 kilometer Fiber with Varying Ambient Temperature*

We also conducted tests with a 1 km fiber, which replaces the 400 m fiber in test scheme B.

The results shown in Fig. 13 indicate that with this new method a clock delay compensation better that 50 ps is achieved with a 1 km fiber.

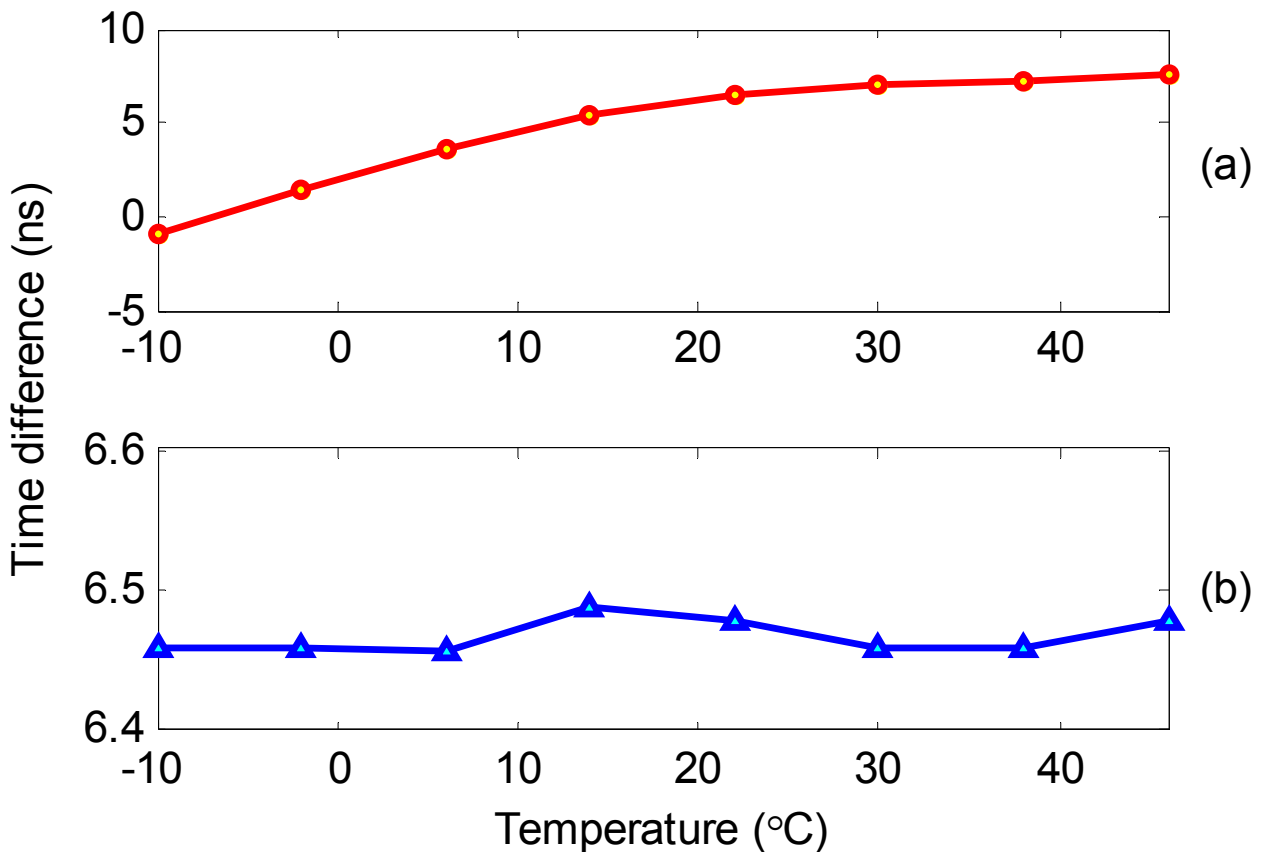

Fig. 13. Test results of scheme C. (a) without compensation; (b) with compensation.

### *D. Jitter Performance*

We also conducted tests to evaluate the clock jitter performance at the FEE (with a clock frequency of 62.5 MHz).

Fig. 14 shows the distribution of the clock period measured at 22°C. The 16 ns mean corresponds to the clock frequency of 62.5 MHz while the rms jitter is 9.2 ps.

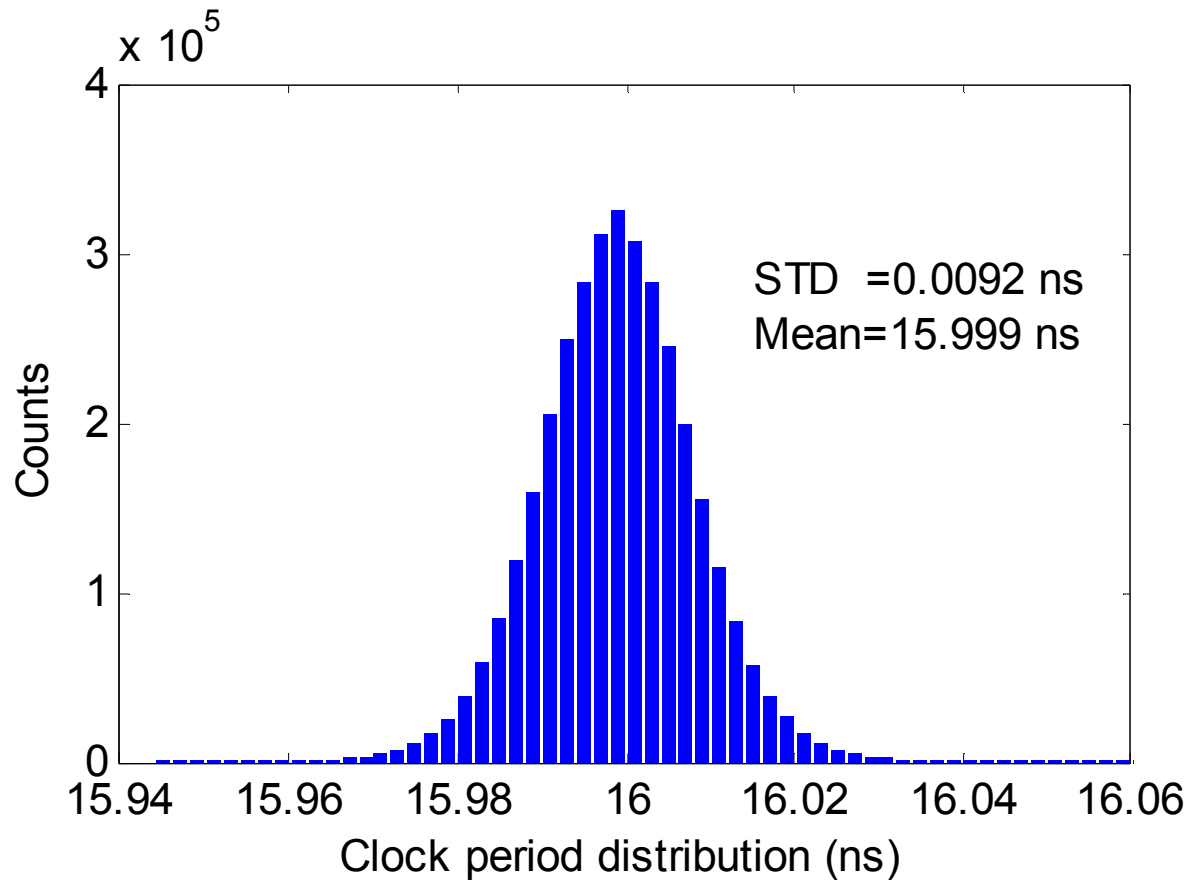

Fig. 14. Clock jitter performance at the FEE at 22 °C.

We also changed the ambient temperature from -10 °C to 46 °C. The test results shown in Fig. 15 indicate that a jitter performance of ~10 ps is successfully achieved with this clock phase compensation method.

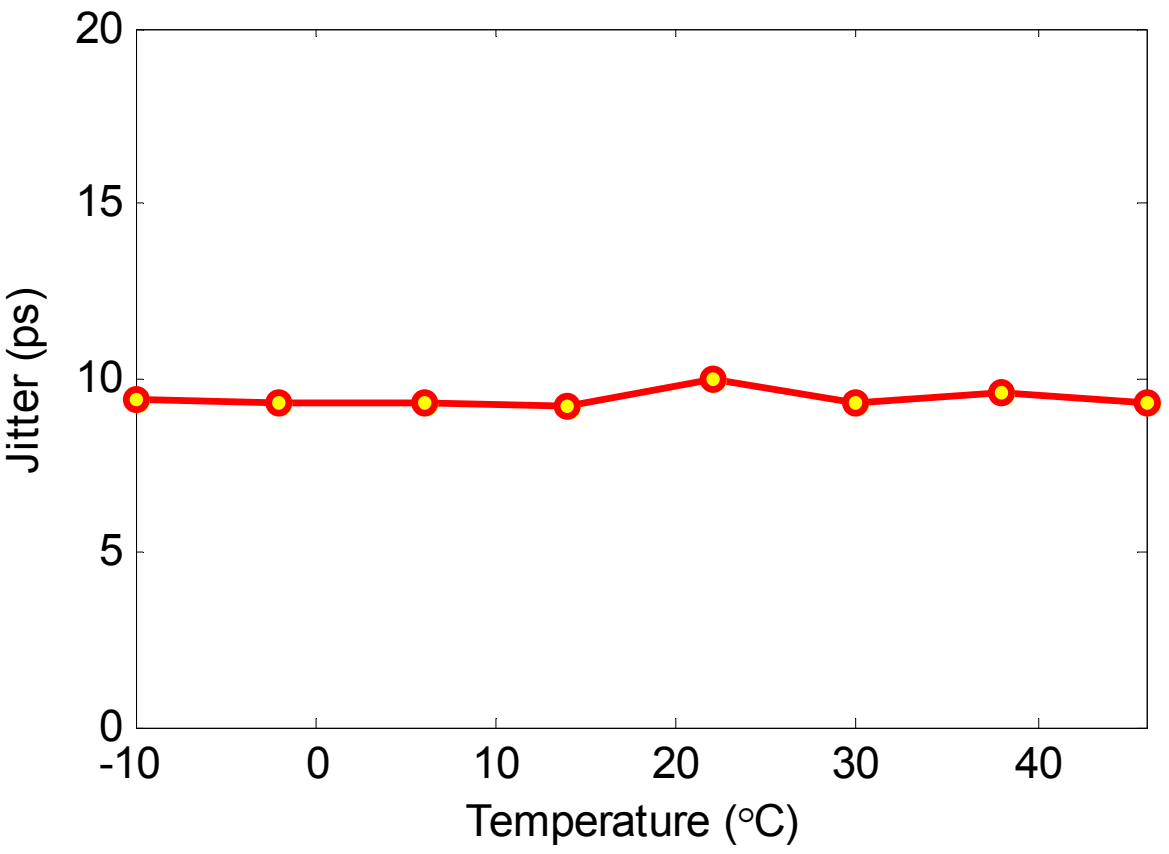

Fig. 15. Clock jitter performance in the temperature range from -10 °C to 46 °C.

## V. Summary and Discussion

According to the above analysis and test results, the clock phase can be automatically compensated among FEEs with the proposed method. In the future design and the installation of the overall 400 FEEs (for the readout of 3600 PMTs), the clock synchronization will consist of two steps. The first step is to calibrate the clock phases of these FEEs at a certain temperature (organized in groups, tested, and compared with a common FEE as the reference), and use the information as the clock phase references for these FEEs. The second step is to use the compensation method proposed in this paper, which is based on the real time division of the increment value of the roundtrip delay compared with the above clock phase reference value.

Offline calibration in the first step can be conducted in a traditional way, while the second step guarantees the clock phase alignment in real time when the temperature changes, without the need of temperature monitoring or complex LUTs.

## VI. Conclusion

By studying the temperature dependence of the fiber and electronics delays, we achieved completely automatic clock synchronization over a distance of up to 1 km based on a new clock phase alignment/compensation method. Combined with the hardware design to achieve a good downwards and upwards circuit symmetry, a good synchronization precision is achieved. Our test results indicate that, over a 56° C temperature variation span, the synchronization variation is below 50 ps (with 1 km fibers) and the jitter performance is ~10 ps rms. These performance figures are well beyond the requirements of our application.